\documentclass[12pt,a4paper]{JHEP}
\usepackage{epsfig,axodraw}
\usepackage{array}
\usepackage{cite}
\usepackage{amsmath}
\usepackage{amssymb}
\usepackage{epic}

\newcommand{\inmath}[1] {\ifmmode#1\else$#1$\fi}
\newcommand{\definmath}[2] {\def#1{\ifmmode#2\else$#2$\fi}}
\definmath{\emin} {\mathrm{e}^{-}}      
\definmath{\eplus} {\mathrm{e}^{+}}      
\definmath{\epm} {\mathrm{e}^{\pm}}      
\definmath{\emp} {\mathrm{e}^{\mp}}      
\definmath{\ee} {\eplus\emin}      
\definmath{\mumin} {\mu^{-}}      
\definmath{\muplus} {\mu^{+}}      
\definmath{\mupm} {\mu^{\pm}}      
\definmath{\mump} {\mu^{\mp}}      
\definmath{\mumu} {\muplus\mumin}      
\definmath{\taumin} {\tau^{-}}      
\definmath{\tauplus} {\tau^{+}}      
\definmath{\taupm} {\tau^{\pm}}      
\definmath{\taump} {\tau^{\mp}}      
\definmath{\tautau} {\tauplus\taumin}      
\definmath{\q}  {\mathrm{q}}         
\definmath{\aq}  {\overline{\mathrm{q}}}         
\definmath{\qq}  {\q\aq}         
\definmath{\bp}  {\mathrm{b}'}         
\definmath{\abp}  {\overline{\mathrm{b}'}}         
\definmath{\bpbp}  {\bp\abp}         
\definmath{\z}  {\mathrm{Z}^{0}}
\definmath{\zgamma}  {\z /\gamma}
\definmath{\w}  {\mathrm{W}}
\definmath{\wmin}  {\mathrm{W}^{-}}
\definmath{\wplus}  {\mathrm{W}^{+}}
\definmath{\wpm}  {\mathrm{W}^{\pm}}
\definmath{\ww}  {\wplus\wmin}
\definmath{\mw}  {\mathrm{M}_{\mathrm{W}}}
\definmath{\alphaem}  {\alpha_{\mathrm{em}}}
\definmath{\s}  {\mathrm{s}}
\definmath{\sp}  {\mathrm{s}'}
\definmath{\roots}  {\sqrt{\s}}
\definmath{\rootsp}  {\sqrt{\sp}}
\definmath{\sigmaqqx}  {\sigma(\qq\mathrm{X})}
\definmath{\dedx}  {de/dx}
\definmath{\mev}  {\mathrm{MeV}}
\definmath{\gev}  {\mathrm{GeV}}
\definmath{\pb}  {\mathrm{pb}}
\definmath{\invpb}  {\mathrm{pb}^{-1}}

\definmath{\nue} {\nu_{e}}
\definmath{\numu} {\nu_{\mu}}
\definmath{\nutau} {\nu_{\tau}}

\definmath{\Ds}  {ds}
\definmath{\Dstar}  {ds}
\definmath{\qqbar}  {qqbar}

\definmath{\gluino} {\tilde{g}}
\definmath{\sqr} {\tilde{q}_{R}}
\definmath{\sql} {\tilde{q}_{L}}

\definmath{\ssul} {\tilde{u}_{L}}
\definmath{\ssdl} {\tilde{d}_{L}}
\definmath{\sscl} {\tilde{c}_{L}}
\definmath{\sssl} {\tilde{s}_{L}}
\definmath{\sstone} {\tilde{t}_{1}}
\definmath{\ssbone} {\tilde{b}_{1}}

\definmath{\ssur} {\tilde{u}_{R}}
\definmath{\ssdr} {\tilde{d}_{R}}
\definmath{\sscr} {\tilde{c}_{R}}
\definmath{\sssr} {\tilde{s}_{R}}
\definmath{\ssttwo} {\tilde{t}_{2}}
\definmath{\ssbtwo} {\tilde{b}_{2}}

\definmath{\ssulbr} {\bar{\tilde{u}_{L}}}
\definmath{\ssdlbr} {\bar{\tilde{d}_{L}}}
\definmath{\ssclbr} {\bar{\tilde{c}_{L}}}
\definmath{\ssslbr} {\bar{\tilde{s}_{L}}}
\definmath{\sstonebr} {\bar{\tilde{t}_{1}}}
\definmath{\ssbonebr} {\bar{\tilde{b}_{1}}}

\definmath{\ssurbr} {\bar{\tilde{u}_{R}}}
\definmath{\ssdrbr} {\bar{\tilde{d}_{R}}}
\definmath{\sscrbr} {\bar{\tilde{c}_{R}}}
\definmath{\sssrbr} {\bar{\tilde{s}_{R}}}
\definmath{\ssttwobr} {\bar{\tilde{t}_{2}}}
\definmath{\ssbtwobr} {\bar{\tilde{b}_{2}}}

\definmath{\chginoonem} {\tilde{\chi}_{1}^{-}}
\definmath{\chginoonep} {\tilde{\chi}_{1}^{+}}
\definmath{\chginoonepm} {\tilde{\chi}_{1}^{\pm}}

\definmath{\chginotwom} {\tilde{\chi}_{2}^{-}}
\definmath{\chginotwop} {\tilde{\chi}_{2}^{+}}
\definmath{\chginotwopm} {\tilde{\chi}_{2}^{\pm}}

\definmath{\chginoall} {\tilde{\chi}_{1,2}^{\pm}}

\definmath{\ntlinoone} {\tilde{\chi}_{1}^{0}}
\definmath{\ntlinotwo} {\tilde{\chi}_{2}^{0}}
\definmath{\ntlinothree} {\tilde{\chi}_{3}^{0}}
\definmath{\ntlinofour} {\tilde{\chi}_{4}^{0}}
\definmath{\ntlinoall} {\tilde{\chi}_{1,2,3,4}^{0}}

\definmath{\ssellp} {\tilde{e}_{L}^{+}}
\definmath{\ssellm} {\tilde{e}_{L}^{-}}
\definmath{\ssellpm} {\tilde{e}_{L}^{\pm}}

\definmath{\sselrp} {\tilde{e}_{R}^{+}}
\definmath{\sselrm} {\tilde{e}_{R}^{-}}
\definmath{\sselrpm} {\tilde{e}_{R}^{\pm}}

\definmath{\ssmulp} {\tilde{\mu}_{L}^{+}}
\definmath{\ssmulm} {\tilde{\mu}_{L}^{-}}
\definmath{\ssmulpm} {\tilde{\mu}_{L}^{\pm}}

\definmath{\ssmurp} {\tilde{\mu}_{R}^{+}}
\definmath{\ssmurm} {\tilde{\mu}_{R}^{-}}
\definmath{\ssmurpm} {\tilde{\mu}_{R}^{\pm}}

\definmath{\sstauonep} {\tilde{\tau}_{1}^{+}}
\definmath{\sstauonem} {\tilde{\tau}_{1}^{-}}
\definmath{\sstauonepm} {\tilde{\tau}_{1}^{\pm}}

\definmath{\sstautwop} {\tilde{\tau}_{2}^{+}}
\definmath{\sstautwom} {\tilde{\tau}_{2}^{-}}
\definmath{\sstautwopm} {\tilde{\tau}_{2}^{\pm}}

\definmath{\sslrpm} {\tilde{l}_{R}^{\pm}}
\definmath{\sslr} {\tilde{l}_{R}}
\definmath{\ssll} {\tilde{l}_{L}}

\definmath{\ssnuel} {\tilde{\nu}_{e}}
\definmath{\ssnumul} {\tilde{\nu}_{\mu}}
\definmath{\ssnutl} {\tilde{\nu}_{\tau}}

\definmath{\M} {\times10^{6}}
\definmath{\K} {\times10^{3}}

\title{Measuring Supersymmetric Particle Masses at the LHC in 
Scenarios with Baryon--Number R-Parity Violating Couplings}

\author{B.C. Allanach\\Theory Division, CERN, 1211 Geneva 23, Switzerland.}
\author{A.J. Barr, L. Drage, C.G. Lester, D. Morgan, M.A. Parker,
	B.R. Webber\\
        Cavendish Laboratory, University of Cambridge, Madingley Road,
        Cambridge, CB3\nolinebreak\ \nolinebreak{0HE,} UK.}

\author{P. Richardson\\
	DAMTP, Centre for Mathematical Sciences, Wilberforce Road, Cambridge,
        \mbox{CB3~0WA,~UK}, and Cavendish Laboratory, University of Cambridge,
	  Madingley Road, 
        Cambridge, CB3\nolinebreak\ \nolinebreak{0HE,} UK.}

\abstract{
	The measurement of sparticle masses in the Minimal Supersymmetric 
	Standard Model at the LHC is analysed, in the scenario where the 
	lightest neutralino, the \ntlinoone, decays into three quarks.
	Such decays, occurring through the baryon-number violating coupling 
	$\lambda''_{ijk}$, pose a severe challenge to the capability of
	the LHC detectors since the final state
	has no missing energy signature and a high jet multiplicity. 
	We focus on the case $\lambda''_{212} \ne 0$ which is the
	most difficult experimentally.
	The proposed method is valid over a wide range of SUGRA
	parameter space with $\lambda''_{212}\sim 10^{-5}-0.1$.
	Simulations are performed of the ATLAS
	detector at the Large Hadron Collider.
	Using the \ntlinoone\ from the decay chain
	$\sql
	\rightarrow\ntlinotwo q
	\rightarrow\sslr\ell q
	\rightarrow\ntlinoone\ell\ell q$,
	we show that the \ntlinoone\ and \ntlinotwo\ masses can be 
	measured by 3-jet and 3-jet + lepton pair invariant mass
	combinations. At the SUGRA point $m_0=100$~GeV, 	
	$m_{1/2}=300$~GeV, $A_0=300$~GeV, $\tan\beta=10$, 
	$\mu>0$
	and with $\lambda^{\prime\prime}_{212}=0.005$,
	we achieve statistical (systematic) errors  of 
	3 (3), 3 (3), 0.3 (4) and 5 (12)
	GeV respectively for the masses of the
	$\ntlinoone$, $\ntlinotwo$, $\sslr$ and $\sql$, with 
	an integrated luminosity of $30~\rm{fb}^{-1}$.
}
\preprint{Cavendish HEP-2001-02\\
	DAMTP-2001-7\\
	ATLAS-COM-PHYS-2001-003\\
	CERN-TH-2001-011}
\begin{document}
\section{Introduction}

The addition of supersymmetry (SUSY) to the Standard Model (SM) represents
a theoretically attractive way of addressing several of the problems faced
when attempting to reconcile constraints from fundamental models at high
scales with the phenomenology seen at the electroweak scale. In
general, supersymmetric models which attempt to solve the naturalness
problem, and fix the Higgs mass near the electroweak scale, predict a rich
spectrum of physics which can be accessed by experiments at the Large
Hadron Collider (LHC). However, most studies of experimental signals for
SUSY have assumed that R-parity ($\mathrm{R_P}$) is conserved. 
$\mathrm{R_P}$ is a multiplicative
quantum number defined as $(-1)^{3\rm{B}+\rm{L}+2\rm{S}}$
where B and L are baryon and
lepton numbers, and S is the spin of the particle. It therefore
distinguishes SM particles ($\mathrm{R_P}=+1$) from their superpartners
($\mathrm{R_P}=-1$).

If $\mathrm{R_P}$ is conserved (RPC models), SUSY particles can only be
created in pairs, and the lightest SUSY particle (LSP) is stable.
Therefore SUSY events each contain an even number of LSPs, which escape
detection and give rise to a large missing transverse energy
($E_T^{\rm{miss}}$). This signature has been exploited by many analyses
of the discovery potential of the LHC 
\cite{phystdr}, since it provides a clean
separation between SUSY events and the SM background. However, the incomplete
measurement of the final state makes the reconstruction of the SUSY mass
spectrum more difficult.

The following terms \cite{dreiner} may be added to the Minimal SUSY
Standard Model (MSSM) superpotential in order to incorporate R-parity
violation (RPV):
\begin{equation}
W_{\not R_{p}} = \frac{1}{2}
\lambda_{ijk}L_{i}L_{j}\bar{E}_{k} + \lambda_{ijk}^{\prime}
L_{i}Q_{j}\bar{D}_{k} + 
\frac{1}{2}\lambda_{ijk}^{\prime\prime}\bar{U_{i}}\bar{D_{j}}\bar{D_{k}}
+\kappa_iL_{i}H_{2},
\label{equ:rpv_terms}
\end{equation}
where gauge indices have been suppressed.

In the MSSM Lagrangian, SM and SUSY particles are grouped together into
lepton ($L_{i}$), quark ($Q_{i}$) and Higgs ($H_{1,2}$) 
SU(2) doublet superfields and electron ($E_i$),
down ($D_i$) and up ($U_i$) SU(2) singlet superfields. In
Equation~\ref{equ:rpv_terms}, $\lambda_{ijk}$, $\lambda^{\prime}_{ijk}$ and
$\lambda^{\prime\prime}_{ijk}$ are Yukawa couplings between the matter
superfields and $\kappa_i$ is the 
mixing term between the lepton and Higgs doublets.
The subscripts $i$, $j$ and $k$ are family
indices. $\lambda_{ijk}$ is antisymmetric under $i\leftrightarrow j$ and
$\lambda^{\prime\prime}_{ijk}$ is antisymmetric under $j
\leftrightarrow k$. $W_{\not R_{p}}$ therefore adds $9+27+9+3=48$ free
parameters to the MSSM superpotential.

Interactions involving the $\lambda_{ijk}$, $\lambda^\prime_{ijk}$ and
$\kappa_i$
couplings violate lepton number, while those involving the
$\lambda^{\prime\prime}_{ijk}$ coupling violate baryon number.
The simultaneous presence of the second and third terms in 
Equation~\ref{equ:rpv_terms} can lead to fast proton decay in gross 
conflict with the lower limit on the proton lifetime \cite{Groom:2000in}.
Since both lepton and baryon number violation 
are required in order for the proton to decay, 
current experimental bounds on the proton lifetime and other SM
processes can be respected if either baryon number or lepton number is
conserved \cite{dreiner}. 

The difference in experimental signatures between RPV and RPC SUSY models
at the LHC depends on the strength of the RPV coupling. 
We will concentrate on the trilinear couplings and neglect the bilinear
term which leads to mixing between the leptons and gauginos, and between
the sleptons and Higgs bosons.
When the RPV couplings
are small compared to the MSSM gauge couplings, the dominant effect is
that the LSP can decay into SM particles. For example, the lifetime of the
LSP ($\ntlinoone$) as a function of the RPV coupling $\lambda''_{212}$ is
shown in Figure~\ref{fig:rpvbranch}a. 

\FIGURE[t]{
\includegraphics[angle=90,width=0.45\textwidth]{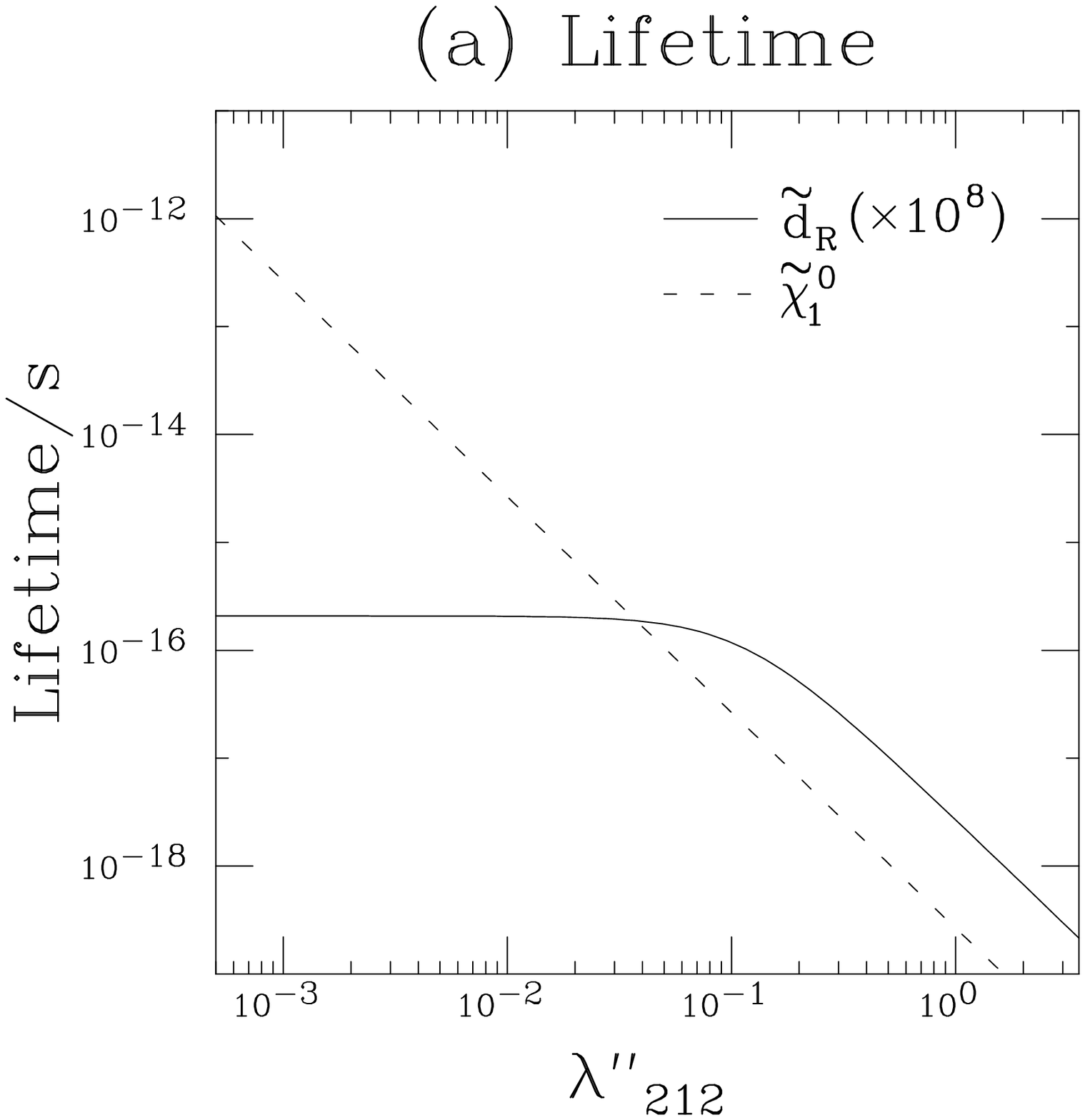}
\hfill
\includegraphics[angle=90,width=0.45\textwidth]{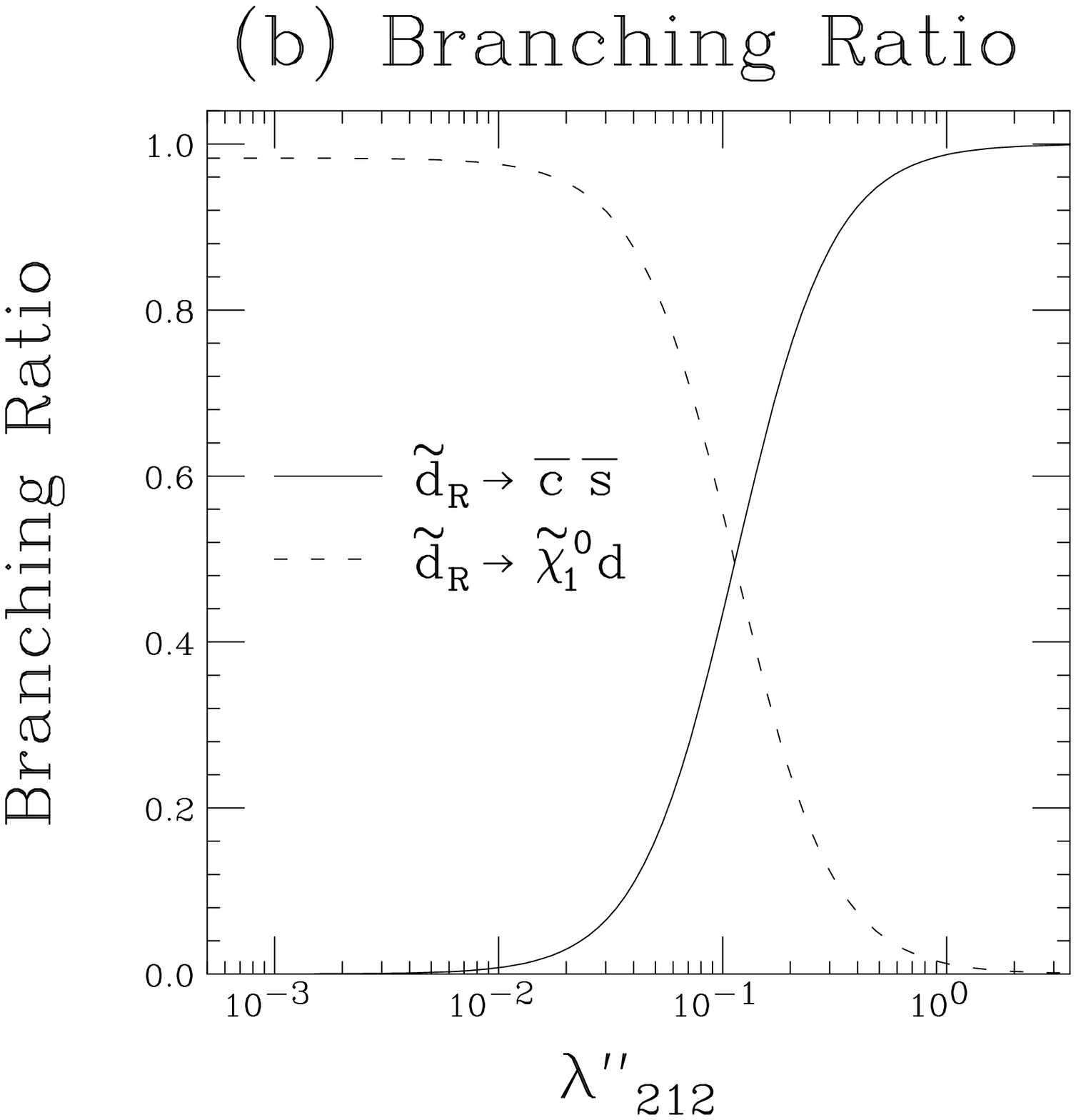}\\
\caption{{\bf (a)} Lifetimes of the $\tilde{d}_R$ and $\ntlinoone$ and
	 {\bf (b)} branching ratio of RPC (dashed) and RPV (solid) decays 
	of the $\tilde{d}_R$,
         plotted
         against $\lambda''_{212}$ at the SUGRA point $m_0=100$~GeV, 
	 $m_{1/2}=300$~GeV, $A_0=300$~GeV, $\tan\beta=10$ and
	 $\rm{sgn}\,\mu\ +$.}
\label{fig:rpvbranch}
}

If the RPV couplings and MSSM gauge couplings are of the same order
of magnitude, RPV production processes and decays of particles
heavier than the LSP become important.  For a large
$\lambda^{\prime\prime}_{ijk}$ coupling, the branching
fractions of RPC and RPV decays of a squark can be of the same order of
magnitude (Figure~\ref{fig:rpvbranch}b), and SUSY particles can be
produced singly, as was investigated in \cite{Chiappetta}.

In the present analysis, $\lambda^{\prime\prime}_{212}$ is the only
RPV coupling set to a non-zero value. This gives rise to the LSP decay
mode $\ntlinoone\rightarrow cds$. This is the most challenging case
experimentally, since there are no leptons or $b$-quarks among 
the $\ntlinoone$ decay products which
can be used as tags for signal events. As each event will
usually contain two
$\ntlinoone$'s there will be at least six jets in the final state. 

The RPV coupling was added to a minimal supergravity (mSUGRA) model with
5 GUT-scale parameters: a universal scalar mass $m_0=100$ GeV, a universal
gaugino mass $m_{1/2}=300$ GeV, trilinear $H\tilde{f}\tilde{f}$ soft SUSY
breaking terms
$A_0=300$ GeV, the ratio of the vacuum expectation values of the two
Higgs doublets $\tan\beta=10$ and the sign of the SUSY
higgsino mass parameter $\mu$ positive. 
It should be noted that we have only included the RPV coupling at the weak
scale, i.e. the RPV coupling is not used in the evolution from the GUT
scale, as was done in \cite{Allanach:1999mh}.

Five sets of parameter values have been extensively studied in the 
$\mathrm{R_P}$ conserving MSSM. The parameters chosen here
correspond to SUGRA Point~5, with one modification: the value of
$\tan\beta$ has been increased from 2.1 to 10 in order to keep the
predicted Higgs mass above the current experimental limit.
The masses of some key particles in this model are given in 
Table~\ref{tab:mass}.
Searches in
the SUGRA Point~5 scenario have been well studied in the case of a stable
$\ntlinoone$
\cite{phystdr}. 
At this SUGRA point, the \ntlinoone\  is the LSP, as must be the case
for our analysis to be valid, even
though cosmological constraints which require the LSP to be neutral only
apply if it is stable.

The
case of $\lambda^{\prime\prime}_{212}=0.005$ is considered first. This
coupling strength gives rise to decay chains essentially identical to
those in an RPC model, except for the decay of the $\ntlinoone$ inside
the beam pipe, with a lifetime of $1.0\times 10^{-14}$~s.
Unlike many other RPV couplings, $\lambda^{\prime\prime}_{212}$ is not
currently constrained by experiment \cite{Allanach}.

\TABULAR{|l|l|l|l|l|l|l|l|l|l|}{
\hline
\ntlinoone &	\ntlinotwo &	\gluino &	\ssur &	\ssul &
	\ssdr &	\ssdl &	\sslr &	\ssll	& $h^0$ \\ 
\hline
116.7 &		211.9 &		706.3 &		611.7 &	632.6 &	610.6 &
	637.5 &	155.3 &	230.5 	& 112.7 \\	
\hline
}{
\label{tab:mass}
Masses of selected particles (GeV) for the model investigated.
}

The effect of varying the RPV coupling in our analysis will be discussed
in Section~\ref{lambda}. For RPV couplings of order
$10^{-6}$ or smaller~\cite{mirea}, the LSP has a sufficiently long
lifetime to decay outside of the detector.
If the LSP is charged, heavily
ionizing low velocity tracks would be seen in the detector, providing a
clear signature. This analysis addresses the case of a neutral LSP, the
lightest neutralino (\ntlinoone ), which has negligible interactions with
the detector. For low RPV couplings, the experimental signature is then
identical to that of an RPC model. However, if RPV couplings are
above~$10^{-4}$ \cite{mirea}, the LSP usually decays in the beam pipe
and the missing energy
signature, seen in RPC models, is not present.

\section{Analysis Strategy}

In this work, HERWIG 6.1 \cite{herwig} is used as the event
generator\footnote {The simulation of RPV events in HERWIG is discussed in 
			\cite{Dreiner:2000qz}.},
and the official ATLAS simulation program ATLFAST \cite{atlfast} is used to
simulate the performance of the ATLAS detector. Since each
$\ntlinoone$ decays to three quarks, and in general the decay chain produces
the
$\ntlinoone$ in association with at least one other quark (typically from
squark decay), the mean jet multiplicity ($N_{\rm{jet}}$) in the signal
events is high.

The principal difficulty in measuring the \ntlinoone\  mass is the
identification of the correct jets from the \ntlinoone\  decay. 
Nearly all right-squarks decay via $\sqr \rightarrow \ntlinoone q
\rightarrow qqqq$ and one might therefore expect
$N_{\rm{jet}}=8$ for \sqr\sqr\  production. Gluon radiation by quarks,
however, raises this to an average of 9.2~jets, in spite of the fact
that the three jets from harder
\ntlinoone s are spatially close together and some merging of jets occurs.
In $\sql\sql$ events $N_{\rm{jet}}=10.7$. The increase with respect to the
right-handed states is due to the difference in couplings to charginos
and neutralinos. Gluinos mostly decay into a squark and a quark and
$\tilde{g}\tilde{g}$ events have a higher value of $N_{\rm{jet}}= 12.8$. A
simple algorithm is used to calculate the jet energies, summing the
energy within a cone of size 0.4 about the jet axis in the $\eta-\phi$
plane, and at least 8 jets with $E_T>25$~GeV are required in signal events.

The analysis proceeds in the following steps:
\begin{itemize}
\item{Cuts are applied to reduce the SM background to below 10\% of the
SUSY signal. These cuts rely on the presence of lepton pairs in the signal
events. Such lepton pairs are produced from the decay chain
\mbox{$\ntlinotwo\rightarrow\sslr\, l \rightarrow ll\ntlinoone$} in most
SUSY models. An analysis based on looking for events with two such
\ntlinotwo\  decays was proposed in \cite{Binetruy:1990wc} but the rate
for events with four leptons is much lower than for events with only
one \ntlinotwo\   decay of this type.}

\item{In each signal event, cuts are made on the jet transverse momenta
($p_T$) in order to preferentially select jets from neutralino decays. }

\item{All possible combinations of three of the selected jets are inspected,
and their invariant mass, $m_{jjj}$, calculated.
Events are retained if two combinations are compatible with the same
candidate $\ntlinoone$ mass.}

\item{One of these three-jet $\ntlinoone$ candidates is combined with an 
opposite sign, same flavour (OSSF) lepton pair.
The invariant mass of this system, 
$m_{jjj\ell\ell}$, is a $\ntlinotwo$ candidate.
A clear peak at the $\ntlinoone, \ntlinotwo$ masses in the
$m_{jjj}$, $m_{jjj\ell\ell}$ plane is then observed.}

\item{The $\sslr$ and the $\sql$ masses are reconstructed using 3-jet
	combinations in the 2-dimensional $\ntlinoone, \ntlinotwo$ mass
	peak.}
\end{itemize}

In the following sections, each of these steps is considered in turn.

\section{Standard Model Background}

The SM background in this model is considered in \cite{phystdr}.
It is shown in \cite{Soderqvist} that the inclusive SUSY signal can be
separated from the SM background by requiring that each event contains:
\begin{itemize}
\item{at least 8 jets with $E_T>25$ GeV;} 
\item{at least one jet with $E_T>100$ GeV;}
\item{transverse sphericity$>0.2$, transverse thrust$<0.9$;} 
\item{$m_{T,\rm{cent}}>1$ TeV, where 
$m_{T,\rm{cent}}=\sum\,p_T^{\rm{jet}} + \sum\,p_T^{\rm{lepton}}$, 
where the sum includes central jets and leptons (i.e. with
pseudorapidity $|\eta|<2$);}
\item{at least two leptons ($e$ or $\mu$) with $p_T>15$ GeV 
	and $|\eta|<2.5$.}
\end{itemize} 

\EPSFIGURE[t]{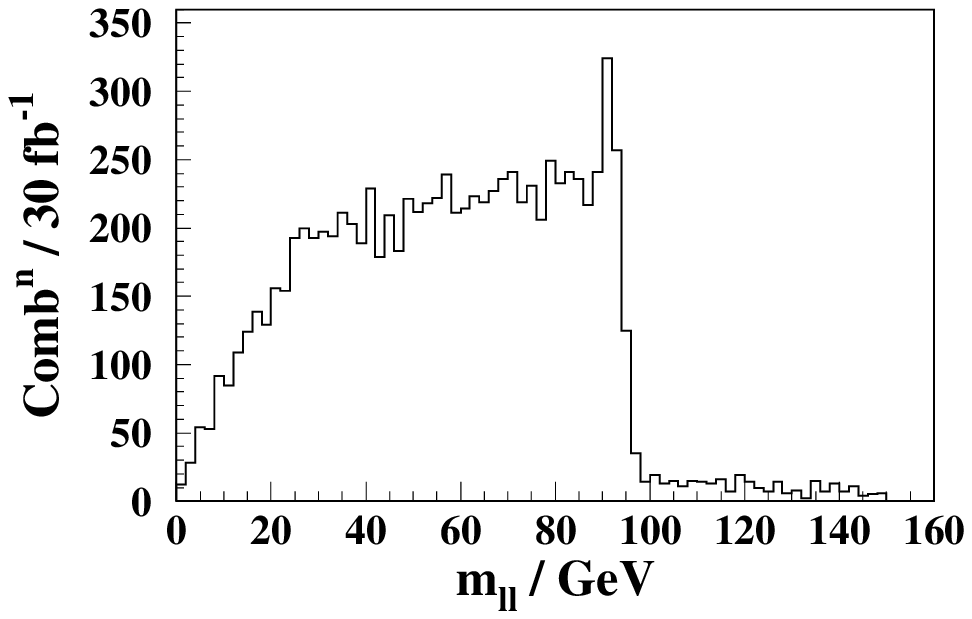, height=7cm}
{The dilepton invariant mass for events with an OSSF electron or muon pair,
after the SM cuts have been applied. The kinematic limit for the decay chain
shown in Figure~\ref{fig:chain} is at 95.1~GeV. Events are excluded if there
is no jet combination which passes the jet cuts described in
Section~\ref{sect:ntl}.
For the SUGRA point chosen it happens that the kinematic edge lies just above
the peak at $m(\z)$ from the decay $\z\to\ell\ell$. 
\label{fig:mll}
}

With these cuts, the signal to background ratio is greater than 10. The
SM background has not been explicitly simulated in this study. Current
Monte Carlo event generators are not capable of reliably simulating QCD eight
jet plus two lepton production. We have therefore simulated eight jets and
two leptons distributed according to phase space.

In SUSY events,
lepton pairs are created in the decay $\ntlinotwo\rightarrow
\tilde{l}_{R}l\rightarrow l l \ntlinoone$. The leptons are therefore required
to have opposite charges and the
 same flavour (OSSF). The invariant-mass distribution of the 
lepton pairs created in this decay has a kinematic edge\cite{phystdr,Lester} 
which is given by 
\begin{equation}
{m}_{\ell\ell}^{\rm max}=\sqrt{\frac
	{[m^2(\ntlinotwo)-m^2(\sslr)]\times[m^2(\sslr)-m^2(\ntlinoone)]}
	{m^2(\sslr)}},
\label{eq:lledge}
\end{equation} 
and is simulated after experimental resolution in Figure~\ref{fig:mll}.

With the particular parameter set chosen this edge is calculated as 95.1~GeV.
Accordingly, events are required to have a lepton pair with an
invariant mass below this value. The corresponding edge can be easily 
measured for other parameter sets.

\section{Detection of the \bf{\ntlinoone} and \bf{\ntlinotwo}}
\label{sect:ntl}

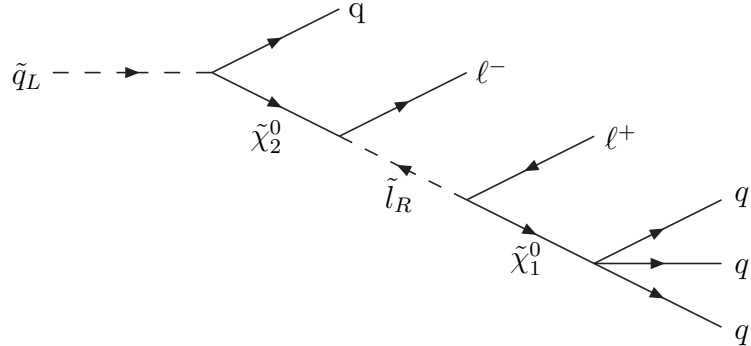
\begin{figure}[htp]
\begin{center}
\begin{picture}(300,110)
\SetScale{1.2}
\SetOffset(20,0)
\DashArrowLine(20,70)(70,70){5}
\ArrowLine(70,70)(110,90)
\ArrowLine(70,70)(110,50)
\ArrowLine(110,50)(150,70)
\DashArrowLine(150,30)(110,50){5}
\ArrowLine(190,50)(150,30)
\ArrowLine(150,30)(190,10)
\ArrowLine(190,10)(230,30)
\ArrowLine(190,10)(230,10)
\ArrowLine(190,10)(230,-10)
\Text(20,83)[r]{\sql}
\Text(136,107)[l]{q}
\Text(105,60)[c]{\ntlinotwo}
\Text(184,85)[l]{$\ell^-$}
\Text(155,38)[c]{\sslr}
\Text(233,60)[l]{$\ell^+$}
\Text(197,14)[l]{\ntlinoone}
\Text(282,37)[l]{$q$}
\Text(282,10)[l]{$q$}
\Text(282,-15)[l]{$q$}
\end{picture}
\end{center}
\caption{One of the decay chains of the \sql\  contributing to the signal.}
\label{fig:chain}
\end{figure}

Many different decay chains can contribute to the SUSY signal selected by
the cuts. One important example, $\sql \rightarrow \ntlinotwo
q \rightarrow \sslr l q \rightarrow \ntlinoone llq \rightarrow qqq llq$,
is shown in Figure~\ref{fig:chain}. 

When $\lambda_{212}^{\prime\prime}$ is small, there are
nearly always two \ntlinoone s produced in an event  and one can therefore
search for two sets of three jets with similar invariant mass. 
An upper limit on the invariant mass difference of
$\delta m_{jjj} = |m^{(a)}_{jjj} -
m^{(b)}_{jjj}|<20$~GeV is used in this analysis, where $a$ and $b$ label the
two $\ntlinoone$ LSP candidates. 

In order to limit the combinatorial background, the search for
the $\ntlinoone$ signal is initially restricted to events with
$8\le N_{\rm{jet}}\le
10$, with the following cuts on the allowed range of jet transverse
momenta in GeV:

\begin{itemize}
\item $100 < p_{T}^{ (a1) }${\bf{;}}
$17.5 < p_{T}^{ (a2) } < 300 ${\bf;}
$15.0 < p_{T}^{ (a3) } < 150 $;
\item $17.5 < p_{T}^{ (b1) } < 300 ${\bf;}
$17.5 < p_{T}^{ (b2) } < 150 ${\bf;}
$15.0 < p_{T}^{ (b3) } < 75 $,
\end{itemize}
where $a1$ denotes the highest $p_{T}$ jet from neutralino candidate $a$,
etc.

\EPSFIGURE[t]{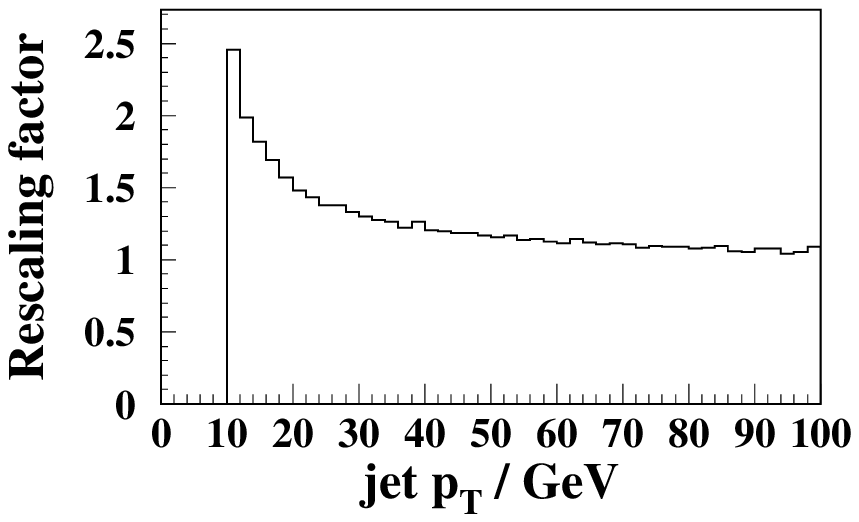, height=8cm}
{The rescaling factor applied to the observed jet energy as a function
of jet $p_{T}$.
\label{fig:calib}
}

\FIGURE[t]{
  \begin{minipage}[b]{.49\linewidth}
    \begin{center}
     \epsfig{file=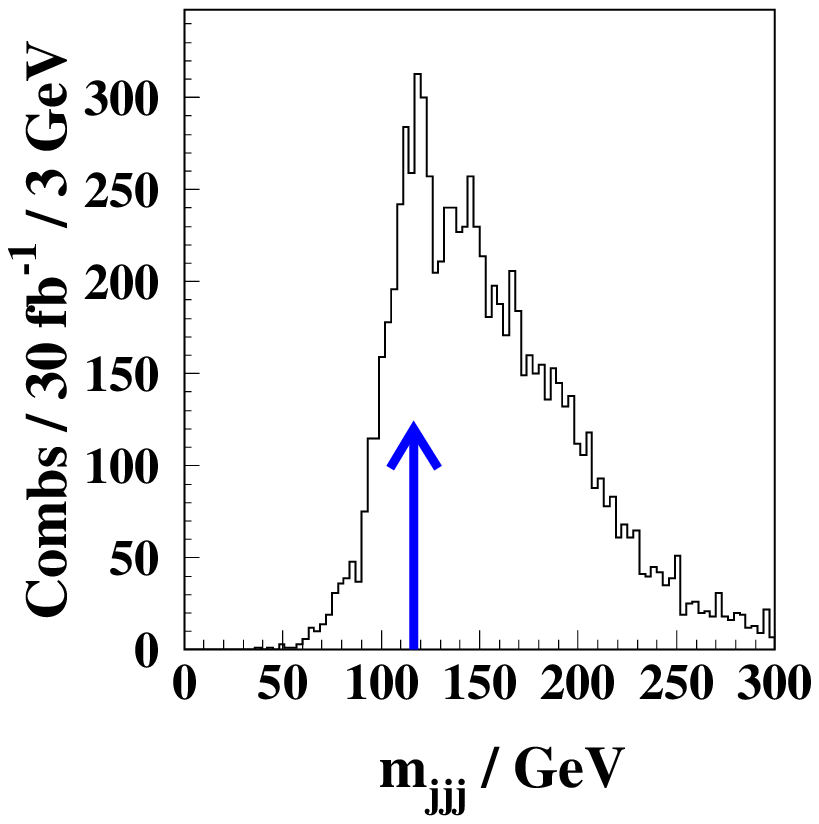, height=7cm,
      width=8.0cm}
      \hspace*{1.2cm}{\bf (a)}
    \end{center}
  \end{minipage}\hfill
  \begin{minipage}[b]{.49\linewidth}
    \begin{center}
     \epsfig{file=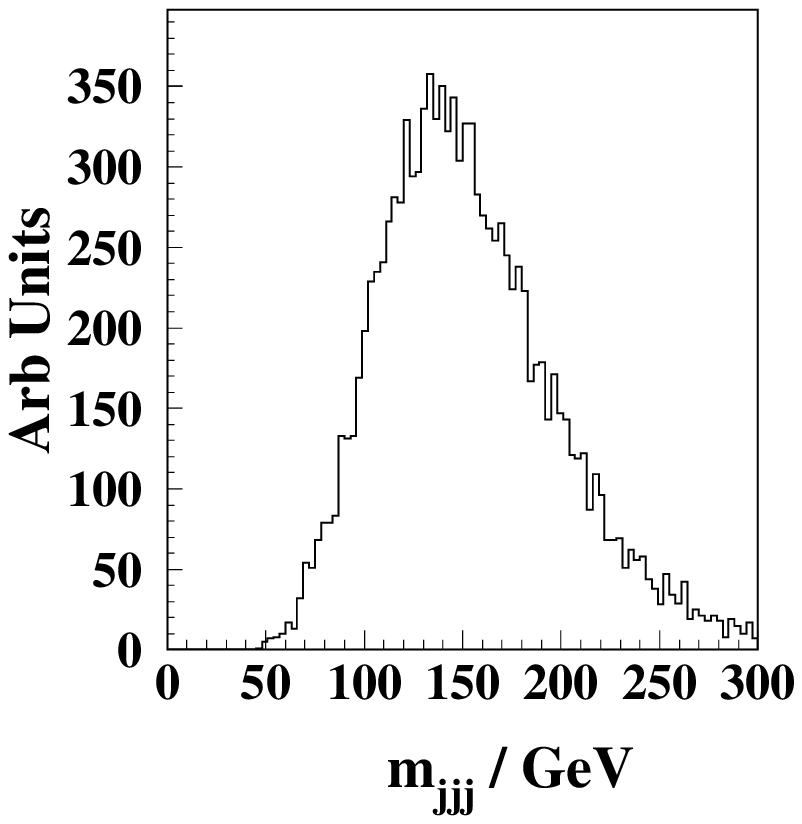, height=7cm,
       width=8.0cm}
       \hspace*{1.4cm}{\bf (b)}
    \end{center}
  \end{minipage}\hfill
\caption{ 
{\bf (a)} The invariant mass of three-jet combinations passing the cuts
described in the text. The mass peak from the decay
\ntlinoone\ $\rightarrow qqq$ can be seen above the background from wrong
combinations of jets. The input \ntlinoone\ mass is indicated by the arrow. 
{\bf (b)} A phase-space sample shows a peak in much the same region.}
\label{fig:jjj} 
}

Candidate sets of jets from the \ntlinoone\  decay can also be identified by
their separation in the $\eta-\phi$ plane. For both \ntlinoone s, cuts
are made on the distance between the hardest and next hardest jets
($\Delta R_{12}$) and between the combined momentum vector of the two
hardest quarks and the softest quark ($\Delta R_{12-3}$). The
following cuts are chosen based on simulations:
\begin{itemize}
\item $\Delta R_{12}^{(a)} < 1.3${\bf;} 
$\Delta R_{12-3}^{(a)} < 1.3${\bf;} 
\item $\Delta R_{12}^{(b)} < 2.0${\bf.}
\end{itemize}

Since the SUSY cross section is dominated by production of squarks and
gluinos, about 95\% of events have two hard jets with
\mbox{$E_T^{(h1)}>200$~GeV} and \mbox{$E_{T}^{(h2)}>100$~GeV} from the
squark decays.
We require that two jets in an event
satisfy these cuts. We do not use those two jets to construct \ntlinoone\
candidates. This decreases the background from wrong combinations.

For each combination of jets passing the kinematic
cuts, the jet energies are rescaled according to their $p_T$, 
to allow for energy lost out of the jet cones. 
The rescaling function, shown in Figure~\ref{fig:calib}, is the ratio
$p_T^{\rm jet}/p_T^{u}$ used in ATLFAST.

The reconstructed masses of all \ntlinoone\ candidates
are included in Figure~\ref{fig:jjj}.
If two or more combinations of jets from one
event pass the cuts, the masses from each combination are plotted with unit
weight. However this can lead to one event
contributing a large number of combinations and therefore if there are more
than five combinations which pass the cuts 
only the five with the smallest difference
between the \ntlinoone\ candidate masses are included.

In events with more than one combination passing the cuts, two combinations
of jets often differ only in the choice of jets for one of the two
\ntlinoone s. In such cases, the unique mass combination is
included in the histogram only once. The two ambiguous masses are both
included with unit weight.

There is a broad combinatorial background
beneath the  \ntlinoone\  mass peak in Figure~\ref{fig:jjj}a, the shape of
which is defined by
the kinematic cuts and is reproduced by the phase-space sample, as shown in
Figure~\ref{fig:jjj}b.
In order to further suppress the background we attempt to find the mass of
the \ntlinotwo\ in the decay chain 
\mbox{$\ntlinotwo\rightarrow\sslr\, l \rightarrow ll\ntlinoone$}
by forming the total invariant mass of the OSSF dilepton pair and one of
the 3-jet candidates.
We choose the \ntlinoone\ candidate which is nearest in $\eta-\phi$ to
either lepton. 

\FIGURE[t]{
  \begin{minipage}[b]{.53\linewidth}
    \begin{center}
     \epsfig{file=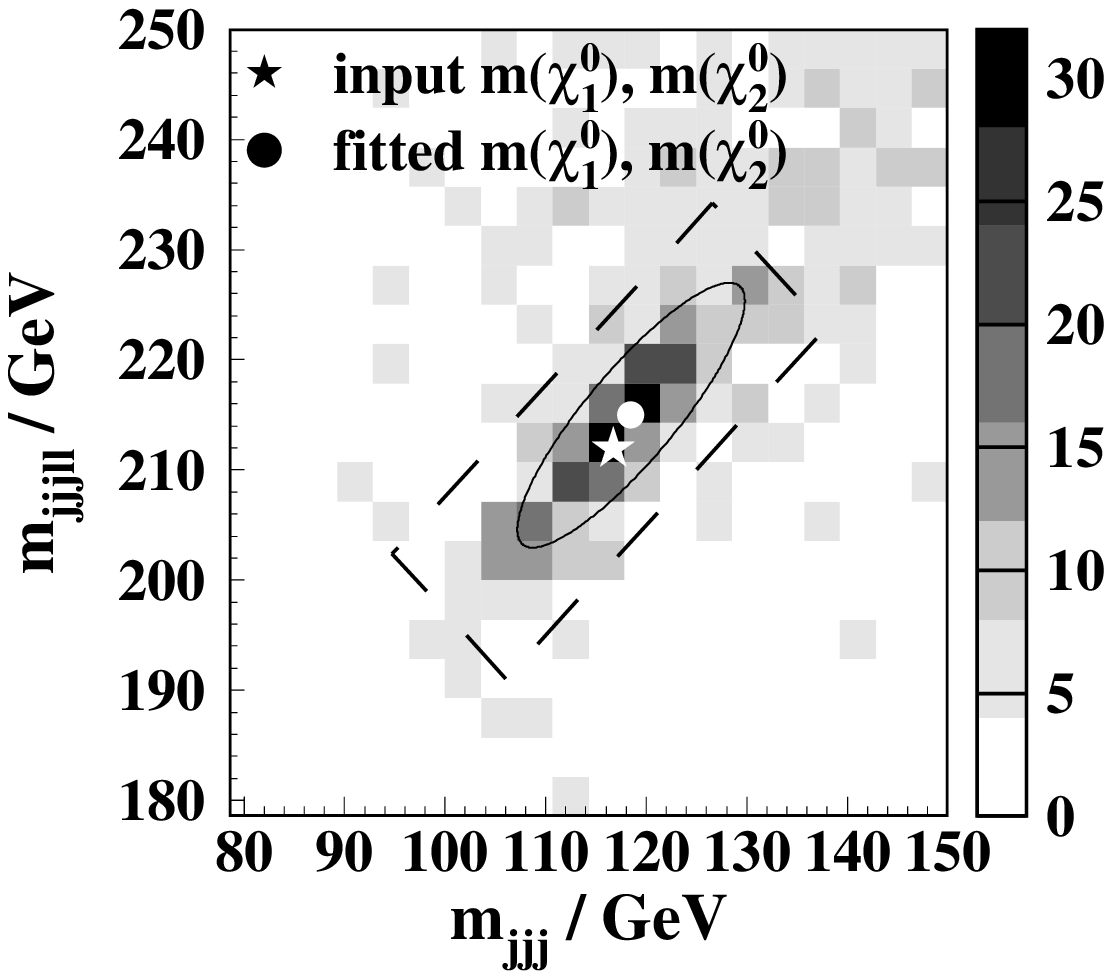, height=7.5cm,
      width=7.7cm}
      {\bf (a)}
    \end{center}
  \end{minipage}\hfill
  \begin{minipage}[b]{.43\linewidth}
    \begin{center}
     \epsfig{file=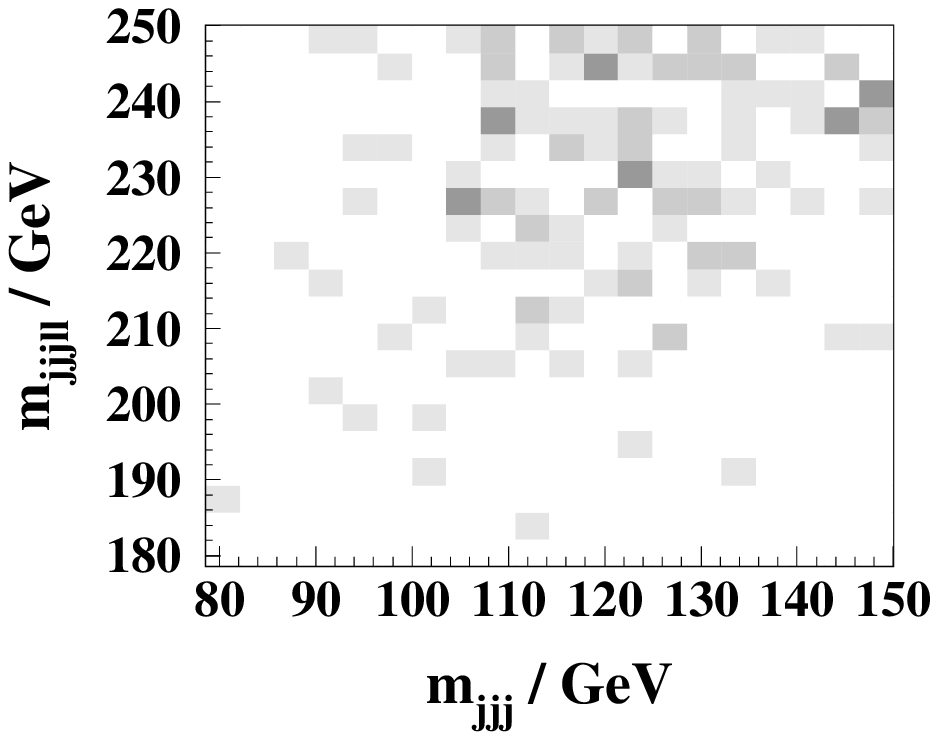, height=7.6cm,
       width=6.5cm}
       \hspace*{1.3cm} {\bf (b)}
    \end{center}
  \end{minipage}\hfill
\caption{ {\bf (a)} The \ntlinoone\ ($m_{jjj}$) and \ntlinotwo\ 
($m_{jjj\ell\ell}$) candidates. The number of jet combinations passing
the cuts per $30~$fb$^{-1}$ is given by the key. The circle and the ellipse
show the peak and standard deviation of a 2-d gaussian fitted to the data
contained in the dashed box. The star shows the input masses.
{\bf (b)} The corresponding ($m_{jjj}$,$m_{jjj\ell\ell}$) invariant mass
combinations from the phase-space sample show no such peak.}
\label{fig:2d_plot} 
}

By using the extra information from the leptons we are able to supress
the combinatorial background. A clear peak in the
(\ntlinoone,\ntlinotwo) mass plane is visible in 
Figure~\ref{fig:2d_plot}a.
Figure~\ref{fig:slices} shows slices through the peak along the axes
$m^{\pm}=(m_{jjj}\pm m_{jjj\ell\ell})/\sqrt{2}$. The peak is present
in the central slices, while the sidebands show similar shapes to the
background under the peak.
It is clear that this peak is not determined by the kinematic cuts,
as it is absent in our phase-space sample (Figure~\ref{fig:2d_plot}b).
In addition we find that the position of the peak accurately follows the
input masses, when those are varied from
(116.7, 211.9) to (137.8,~252.6)~GeV.

The data in the rectangle shown in Figure~\ref{fig:2d_plot} were
fitted with a 2-d gaussian. 
Since $m_{jjj}$ and $m_{jjj\ell\ell}$ are highly correlated, 
the peak was fitted in the rotated~($m^+,m^-$)
coordinate system in which the correlations are smaller. 
The mass difference relies on lepton rather than jet momenta, so the width
in the $m^-$ direction (4~GeV) is smaller than in the $m^+$ direction
(15~GeV).
The standard error on the peak was 4.5 GeV in $m^-$ and 1.5 GeV in $m^+$. 
This corresponds to a 3~GeV uncertainty in each of the neutralino masses.

\EPSFIGURE[t]{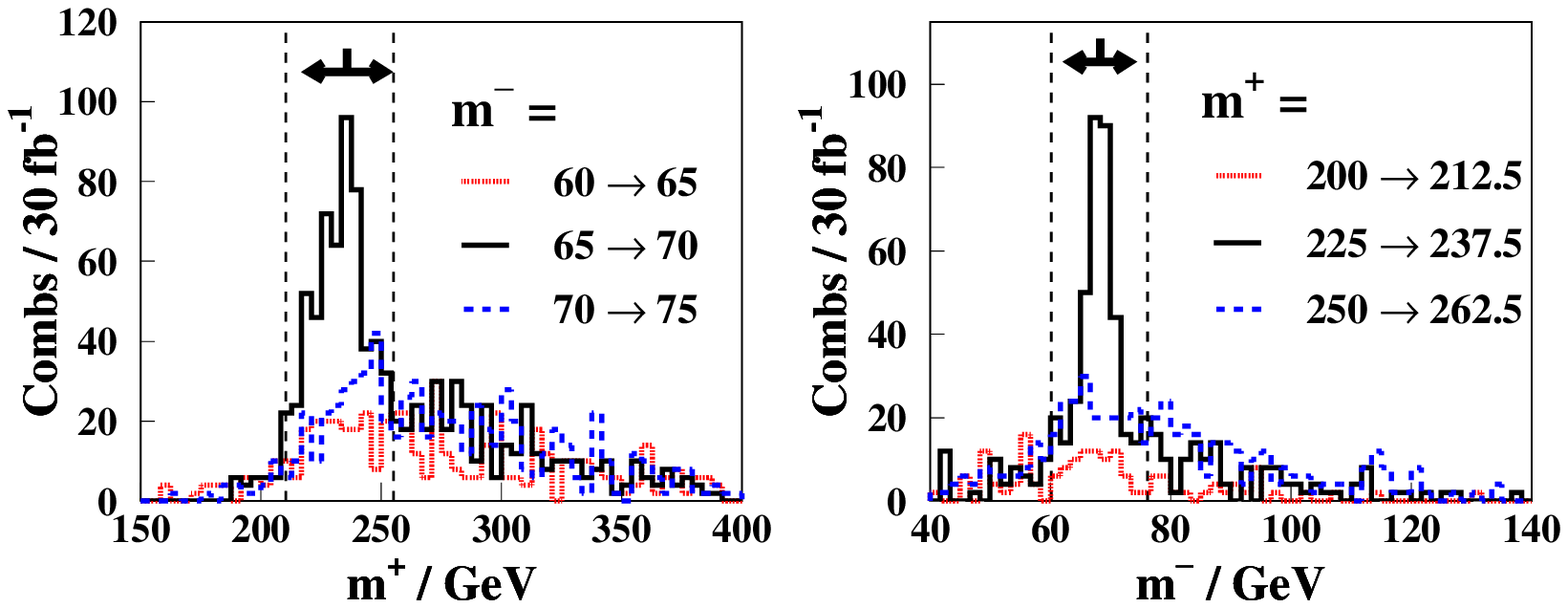, width=\textwidth}
{Slices taken through Figure~\ref{fig:2d_plot}a in the rotated
($m^+$, $m^-$) 
coordinate directions. The fit was performed on the data contained 
in the region between the dotted lines. The fitted peak and width 
are given by the thick line and arrow at the top of each plot.
\label{fig:slices}
}

Our fitted masses, at $m(\ntlinoone,\ntlinotwo)$~=~(118.9, 215.5)~GeV 
are slightly high when compared to the input values of (116.7,~211.9). 
This is due to several effects, including overlap between the jets, 
and the contribution of energy from the underlying event in the jet cones. 
Indeed we would expect that our simple rescaling factor 
will overcompensate for energy losses 
from jet cones, since the three jets from the \ntlinoone\ decay 
are close in $\eta-\phi$, so energy losses from 
one cone can end up in one of the other two.

A fuller investigation of these effects is beyond the scope of this study,
requiring as it does a full investigation of the calorimeter
calibration procedure for multi-jet events.
It is estimated that in the actual experiment, the uncertainty in the 
absolute jet energy scale will be of the order of 1\% for jets 
with $p_T>50$~GeV\cite{phystdr}. For lower energy jets an uncertainty of
2-3\%
is more likely. With real data, therefore, it may be possible to reduce the 
systematic uncertainty on the $\ntlinoone$ mass to 3~GeV.

\section{Detection of the \sslr\ and \sql}

For the measurement of the slepton mass, having fitted 
the \ntlinoone\  and \ntlinotwo\  masses, we select combinations within 
$1\times \sigma$ of the peak. These combinations, with two OSSF leptons 
preferentially select the decay chains 
$\ntlinotwo\rightarrow\sslr l\rightarrow ll\ntlinoone$, and 
$\ntlinotwo\rightarrow\z\ntlinoone$. 
Only the former can be used for \sslr\ measurements. The region of 
mSUGRA parameter space in which this chain will exist is given by the
condition
$m(\ntlinotwo)>m(\sslr)$, and is shown in Figure~\ref{fig:param}.

\EPSFIGURE[t]{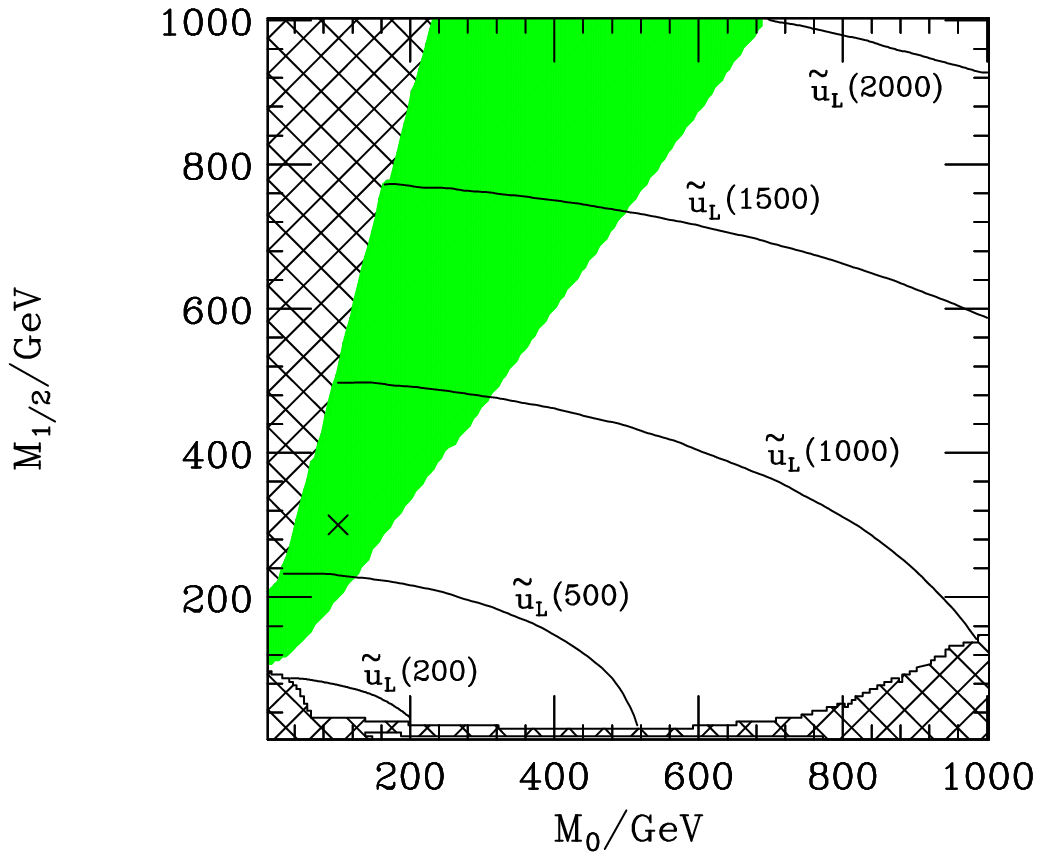, height=8cm}
{The region of mSUGRA parameter space over which the decay 
$\ntlinotwo\rightarrow\sslr\ell \rightarrow \ntlinoone\ell\ell$ occurs is
shaded.
$M_0$ is the universal scalar mass, and $M_{1/2}$ is the universal gaugino
mass at the GUT scale.
The lower hatched region is excluded by lack of electroweak symmetry breaking
or the existence of tachyonic particles, whereas the in the upper
hatched region the \ntlinoone\ is not the LSP and our analysis does not apply.
The contours show the mass of the $\ssul$ squark.
The $\times$ marker shows the chosen point. 
The other parameters are: $\tan\beta=10$, $A_0=300$~GeV and
${\rm sgn}(\mu)=+$.
\label{fig:param}
}

The dilepton invariant mass distribution, Figure~\ref{fig:mll}, for our
point 
in MSSM parameter space shows only a very small peak at the \z\ mass, 
but a clear kinematic edge, indicating that the slepton decay chain
$\ntlinotwo\rightarrow\sslr l\rightarrow ll\ntlinoone$   
dominates. This is expected at this point, 
since the $m(\ntlinotwo)-m(\ntlinoone)$ 
mass difference of 95.2 GeV means that there is 
little phase space available for the decay
$\ntlinotwo\rightarrow\z\ntlinoone$.
If the masses and couplings for this decay were significant, then
we would exclude events with dilepton invariant mass near $m(\z)$ 
from the \sslr\ measurement.
Both chains may have been preceded by $\sql\rightarrow\ntlinotwo q$, 
so both can be used for the \sql\ mass measurement.

\FIGURE[t]{
  \begin{minipage}[b]{.49\linewidth}
    \begin{center}
     \epsfig{file=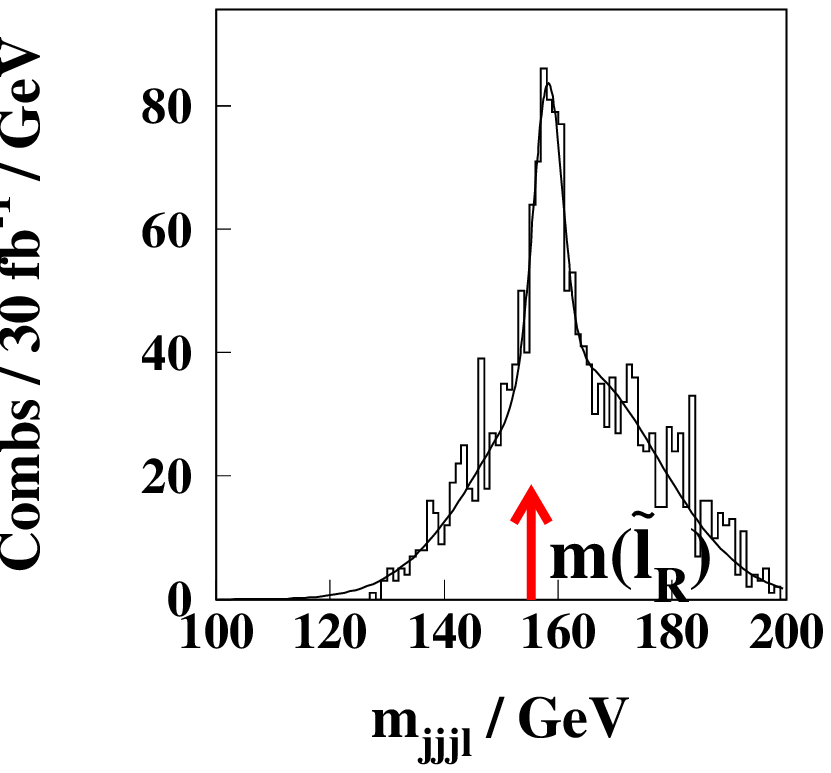, 
	height=8cm}
      {\bf (a)}
    \end{center}
  \end{minipage}\hfill
  \begin{minipage}[b]{.49\linewidth}
    \begin{center}
     \epsfig{file=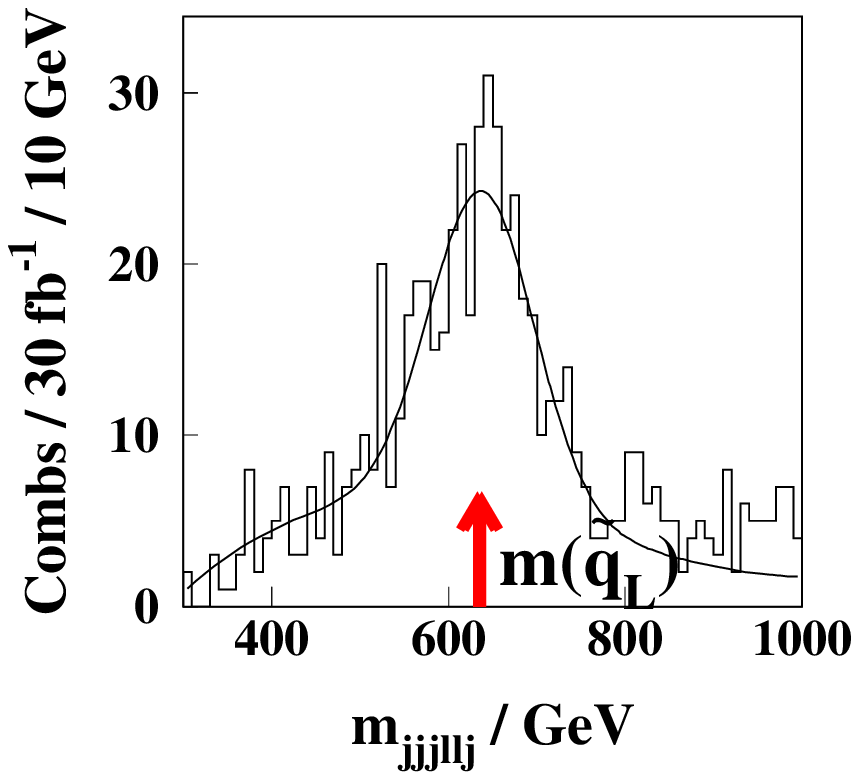, height=8cm}
       {\bf (b)}
    \end{center}
  \end{minipage}\hfill
\caption{ The masses of {\bf (a)} \sslr\ and {\bf (b)} \sql\ candidates.
 	Only combinations from under the (\ntlinoone,\ntlinotwo) mass 
	peak, and satisfying the cuts described in the text are plotted.
	 Each $jjj$ invariant mass has been rescaled to the fitted
	 \ntlinoone\ mass. The \sslr\ and \sql\ masses are indicated
	 by the arrows. The fitted functions are described in the text.}
\label{fig:sleptonmass}
}

We recall that the initial sample of \ntlinoone\  candidates was restricted
to
events with $8\le N_{\rm{jet}}\le 10$ in order to reduce the combinatoric
background. 
However, choosing combinations from under the (\ntlinoone,\ntlinotwo) mass 
peak removes much of the background, so in this section the 
jet multiplicity cut is therefore relaxed to 
$8\le N_{\rm{jet}}\le 11$ in order to increase the statistics. 
The invariant mass of the three-jet neutralino candidates 
is adjusted to the best-fit mass of the \ntlinoone, 
by rescaling the \ntlinoone\ jet momenta by the same factor.

The \sslr\ mass is found by combining the \ntlinoone\  candidate closest
to a lepton in $\eta-\phi$ with that lepton.
The resulting invariant mass
distribution,
$m_{jjjl}$, is shown in Figure~\ref{fig:sleptonmass}a. 
The sharp peak was fitted with a gaussian, 
with another gaussian for the background. This gave
$m(\sslr)=157.8\pm0.3$~GeV, which is slightly high when compared to the 
input value of 155.3~GeV for the same reasons as were discussed in 
Section \ref{sect:ntl}.

The experimental electron and muon momentum scale 
uncertainties are expected to be small fractions of 1\%
\cite{phystdr}, so the systematic error in the slepton mass measurement
will be dominated by the same (3~GeV) jet scale uncertainty as
$m(\ntlinoone)$.
The statistical error in rescaling the 3-jet invariant mass to the fitted 
$m(\ntlinoone)$ peak introduces another 3~GeV systematic error into the 
\sslr\ and \sqr\ masses. The overall systematic error in $m(\sslr)$ is
therefore $3\oplus 3 = 4.2\ \rm{GeV}$.

The hardest two jets in the event are assumed to have come from squark or
gluino 
decay, and so are not used in making three-jet combinations for neutralino
candidates.
The \sql\ mass is found by combining each \ntlinotwo\ candidate with the
harder of these two leading jets.
To reduce the background we select combinations within $2\times \sigma$ of
the m(\sslr) peak. This allows us to relax the cut about the  
(\ntlinoone,\ntlinotwo) peak from 1 to $2\times\sigma$. 
The resultant invariant mass distribution, 
$m_{jjj\ell\ell j}$, is shown in Figure~\ref{fig:sleptonmass}b.
A peak is visible near the \ssul\ and \ssdl\ masses of 
633~GeV and 638~GeV respectively, but the resolution is not sufficient to 
separate the states. 

The background was modelled by finding the invariant-mass
distribution of the 
\ntlinotwo\ candidates with the hardest jet from other \sql\ candidate
events.
A gaussian fit to the signal with this background shape gave
$m(\sql)=637\pm5$~GeV.
The uncertainty in modelling the background was estimated by fitting the
distribution
with other, simpler background shapes. These decrease position of the
peak to
634~GeV (for a resonance-shaped background) to 627~GeV (for a linear
background).
This shows a systematic uncertainty in the \sql\ mass of about 10~GeV.
The hard jet used in the calculation of $m(\sql)$, has $p_T>100\ \rm{GeV}$ 
introducing an uncertainty in the mass scale of 1\%\cite{phystdr}, or 6~GeV.
Carrying forward a 3~GeV uncertainty in the $jjj$ invariant mass scale 
and 3~GeV from the \ntlinoone\ fit, the total 
systematic error in the squark mass is $12$~GeV.

At this SUGRA point the dominant decay mode of the \sqr\  is 
$\sqr\rightarrow\ntlinoone q$. One might therefore try to reconstruct
the \sqr\  mass by combining the
\ntlinoone\  candidate not used in the \sql\  reconstruction with the 
second hard jet. However while the cuts we applied to reduce the
Standard Model background, i.e. requiring the presence of two leptons,
mean that all the signal events contain a \sql,  they do not necessarily
contain a right squark.
Only a third of the signal events actually contain a \sqr.
Most of the SUSY events at this SUGRA point come from gluino production 
which will either be rejected due to the large number of jets or
contain additional hard jets from the gluino decay, and hence have a large
combinatoric background for the \sqr\  reconstruction. In an attempt to
reduce this background it is possible to use cuts on the \ntlinotwo\  
and \sslr\  masses from \sql\  decay on the other side of the event
such that there is only one \sqr\  candidate.
However this reduces the statistics so that a signal cannot be observed.
This combination of factors makes it impossible 
to reconstruct the \sqr\  mass 
at this SUGRA point with the available statistics.

\section{Other values of $\lambda^{\prime\prime}_{212}$}
\label{lambda}

The method outlined above is relatively 
insensitive to the size of the coupling
 $\lambda''_{212}$.
However as the RPV coupling $\lambda^{\prime\prime}_{212}$ is decreased, 
the lifetime of the \ntlinoone\ increases as shown in 
Figure~\ref{fig:rpvbranch}a. The method will start to fail when 
\ntlinoone s predominately decay beyond the first tracking layer 
of the detector. Special reconstruction could increase this by 
about an order of magnitude, at which point the decays would occur 
outside of the tracking volume. We therefore exclude events 
when one or other of the \ntlinoone s has travelled more than
 100~mm (1000~mm)
from the interaction point in the transverse direction.

As can be seen in Figure~\ref{fig:limits} statistics become limiting for
$\lambda''_{212}$ less than about $10^{-5}$, when 
$c\tau\approx 800$~mm.  
With smaller couplings the RPV decay of the \ntlinoone\ 
effectively switches off, and a RPC 
analysis based on a missing transverse energy +~lepton(s) signature, 
such as \cite{Lester}, is effective.

If $\lambda''_{212}$ is larger than 0.1, an initial 
\sqr\ often decays immediately into 2 jets and then only 
one 3 jet invariant mass combination will necessarily be
close to the neutralino mass.

The size of the RPV coupling can be determined from the \ntlinoone\ lifetime, 
as shown in Figure~\ref{fig:rpvbranch}a.
The lifetime is, in principle, measurable for a wide range of couplings,
by using vertexing information. 
However the need for detector-level Monte-Carlo simulation 
makes the measurement of $\lambda''_{212}$ beyond the scope of this paper.

\EPSFIGURE[t]{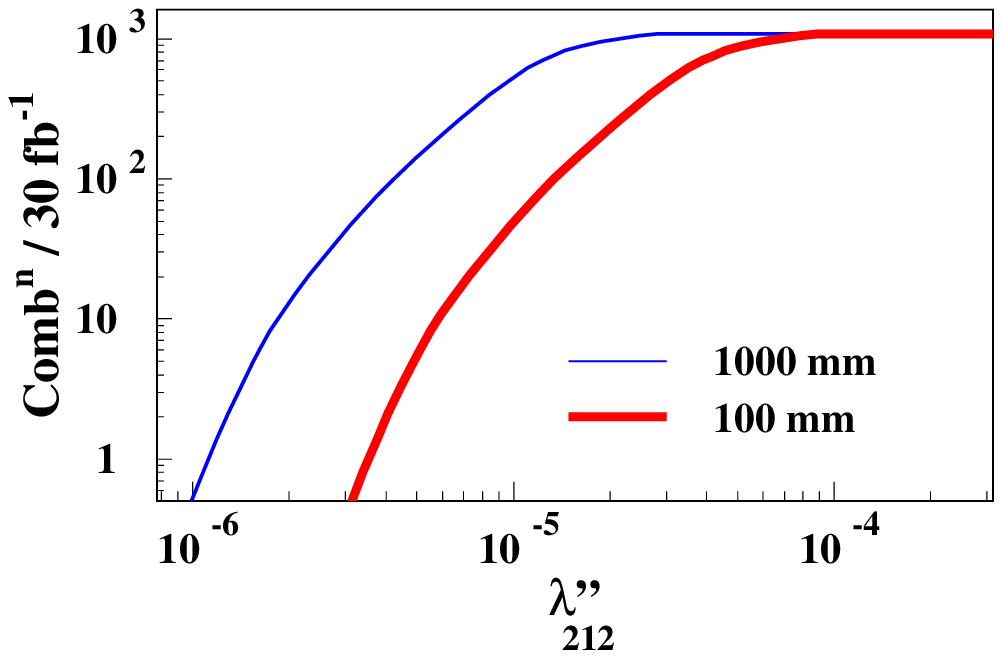,height=7.5cm}
{The number of jet combinations with transverse decay lengths 
less than 100~mm (thick line), and less than 1000~mm (thin line), 
as a function of the RPV coupling.
Combinations beyond $1\times\sigma$ of the (\ntlinoone,\ntlinotwo) 
peak found for $\lambda''_{212}=0.005$ are excluded.
\label{fig:limits}
}

\section{Conclusions}
\label{conclusions}
The usual ubiquitous signature for the $\mathrm{R_P}$ conserving
MSSM is missing transverse energy. This signature disappears 
once $\mathrm{R_P}$ violating couplings are added, 
unless they are extremely small and the lifetime of a neutralino
LSP is such that it decays outside the detector. 

We examined the case where the neutralino LSP 
is unstable and decays to 3 jets with no particular tags on them 
(for example \emph{b}'s). This corresponds to the RPV 
coupling $\lambda''_{212}$, the trilinear RPV coupling giving the 
hardest case in which to detect and measure sparticles. 

By analysing the decay chain, 
	$\sql
	\rightarrow\ntlinotwo q
	\rightarrow\sslr\ell q
	\rightarrow\ntlinoone\ell\ell q$, 
we have shown that the
\ntlinoone, \ntlinotwo\ and \sql\ be detected and their masses measured,
and that the mass of the \sslr\ can also be obtained in much of parameter
space. The sparticle production and decays in this signal 
are all $\mathrm{R_P}$ conserving apart from the \ntlinoone\ decay into
3 jets.

Although we have used a point near SUGRA
Point~5 to derive the soft SUSY breaking parameters, the method should in
principle work for other more general SUSY breaking parameter sets in which
the decay chain in Figure~\ref{fig:chain} exists. 
When some of the sparticles involved in the decay chain become
much heavier than 1~TeV, the analysis will become statistics limited.

To summarise, in the MSSM with a trilinear RPV coupling, even in the
hardest choice of $\lambda''_{212}$, it is possible to detect
sparticles and measure their masses at the LHC.

\acknowledgments

We thank H.~Dreiner and ATLAS collaboration members  for helpful discussions.
We have made use of the physics analysis framework and tools which are
the result of collaboration-wide efforts. This work was partly funded
by PPARC.

\end{document}